\documentclass[10pt,aps,prx,twocolumn,notitlepage,showpacs,superscriptaddress, nofootinbib]{revtex4-1}
\usepackage{amsthm,amsmath,amsfonts,amssymb,verbatim,color}
\usepackage{graphicx}
\usepackage{subfigure}
\usepackage{bm}
\usepackage{epsfig,slashed}
\usepackage[T1]{fontenc}
\usepackage[colorlinks=true,citecolor=blue,linkcolor=blue,urlcolor=blue]{hyperref}
\usepackage{psfrag}
\begin{document}

\newcommand{\tr}{\mathop{\mathrm{Tr}}}
\newcommand{\bsigma}{\boldsymbol{\sigma}}
\newcommand{\re}{\mathop{\mathrm{Re}}}
\newcommand{\im}{\mathop{\mathrm{Im}}}
\renewcommand{\b}[1]{{\boldsymbol{#1}}}
\newcommand{\diag}{\mathrm{diag}}
\newcommand{\sign}{\mathrm{sign}}
\newcommand{\sgn}{\mathop{\mathrm{sgn}}}
\renewcommand{\c}[1]{\mathcal{#1}}
\renewcommand{\d}{\text{\dj}}

\newcommand{\mb}{\bm}
\newcommand{\ua}{\uparrow}
\newcommand{\da}{\downarrow}
\newcommand{\ra}{\rightarrow}
\newcommand{\la}{\leftarrow}
\newcommand{\mc}{\mathcal}
\newcommand{\bs}{\boldsymbol}
\newcommand{\lra}{\leftrightarrow}
\newcommand{\nn}{\nonumber}
\newcommand{\half}{{\textstyle{\frac{1}{2}}}}
\newcommand{\mf}{\mathfrak}
\newcommand{\MF}{\text{MF}}
\newcommand{\IR}{\text{IR}}
\newcommand{\UV}{\text{UV}}
\newcommand{\be}{\begin{equation}}
\newcommand{\ee}{\end{equation}}

\DeclareGraphicsExtensions{.png}

\title{Universality of low-energy Rashba scattering}

\author{Joel Hutchinson}
\affiliation{Department of Physics, University of Alberta, Edmonton, Alberta T6G 2E1, Canada}

\author{Joseph Maciejko}
\email[electronic address: ]{maciejko@ualberta.ca}
\affiliation{Department of Physics, University of Alberta, Edmonton, Alberta T6G 2E1, Canada}
\affiliation{Theoretical Physics Institute, University of Alberta, Edmonton, Alberta T6G 2E1, Canada}
\affiliation{Canadian Institute for Advanced Research, Toronto, Ontario M5G 1Z8, Canada}

\date\today

\begin{abstract}
We investigate the scattering of a quantum particle with a two-dimensional (2D) Rashba spin-orbit coupled dispersion off of circularly symmetric potentials. As the energy of the particle approaches the bottom of the lowest spin-split band, i.e., the van Hove singularity, earlier work has shown that scattering off of an infinite circular barrier exhibits a number of features unusual from the point of view of conventional 2D scattering theory: the low-energy $S$-matrix is independent of the range of the potential, all partial waves contribute equally, the differential cross section becomes increasingly anisotropic and 1D-like, and the total cross section exhibits quantized plateaus. Via a nonperturbative determination of the $T$-matrix and an optical theorem which we prove here, we show that this behavior is universal for Rashba scattering off of any circularly symmetric, spin independent, finite-range potential. This is relevant both for impurity scattering in the noninteracting limit as well as for short-range two-particle scattering in the interacting problem.
\end{abstract}


\maketitle

\section{Introduction}

Long considered a small relativistic correction of little qualitative importance to condensed matter physics, spin-orbit coupling has come to the fore of this field in the past ten years or so owing to the discovery of a rich phenomenology associated with it, including the spin Hall~\cite{sinova2015} and quantum spin Hall~\cite{maciejko2011} effects, three-dimensional (3D) topological insulators~\cite{hasan2010,qi2011}, Weyl semimetals~\cite{armitage2017}, and spin-orbit coupled Mott insulators~\cite{witczak-krempa2014}, to name a few. In 2D crystals with broken inversion symmetry, the spin degeneracy of the electronic band structure may be lifted by Rashba spin-orbit coupling~\cite{rashba1960,bychkov1984}. A similar type of spin-orbit coupling can also be engineering synthetically via laser-atom interactions, as recently demonstrated in an ultracold gas of $^{40}$K fermionic atoms~\cite{huang2016}.

The spin-split dispersion in a 2D Rashba system is described in terms of two distinct helicity bands, but below a threshold energy (Dirac point), particles are confined to one of these. At the bottom of this lower band, the density of states is enhanced to form a van Hove singularity. In particular, this is the relevant regime for a dilute spin-orbit coupled 2D electron gas, which has been shown to host a variety of exotic phases in the presence of electron-electron interactions~\cite{berg2012,silvestrov2014,ruhman2014,bahri2015}. We showed in earlier work~\cite{Hutchinson2016} that in this limit, single-particle scattering from a hard disk potential (i.e., an infinite circular barrier) exhibits a variety of unusual behaviors. The $S$-matrix, which is a $2\times 2$ matrix as there are two same-helicity degenerate scattering channels below the Dirac point, was found in the low-energy limit to be purely off-diagonal with off-diagonal elements equal to $-1$~\cite{hutchinson2017} for every partial wave $l=0,\pm 1,\pm 2,\ldots$ As a result, the low-energy differential scattering cross section is extremely anisotropic, with scattering at all angles $\theta$ highly suppressed except forward scattering ($\theta=0$) and backscattering ($\theta=\pi$), which have the same amplitude. This stands in stark contrast with the usual dominance of the isotropic $s$-wave scattering channel at low energies in conventional systems in both 2D and 3D. Finally, instead of the usual smooth $\sim 1/\sqrt{E}$ divergence (moderated by a logarithm) of the total cross section as the energy $E\rightarrow 0$ in conventional 2D systems with a parabolic dispersion~\cite{friedrich2013}, in the Rashba hard-disk problem the total cross section (which in 2D has units of length) was found to increase in quantized steps of magnitude $4/k_0$ where $k_0$ is the wavenumber of the degenerate Rashba ground state manifold. Surprisingly, all these features are independent of the radius of the barrier, and were found to hold also for a delta-shell potential of arbitrary radius. This led us to conjecture in Ref.~\cite{Hutchinson2016} that these peculiar features are a universal property of low-energy scattering in the Rashba system and should hold for arbitrary spin-independent, circularly symmetric, finite-range potentials.

In the present work we establish that this conjecture is true via a nonperturbative solution of the Lippmann-Schwinger equation for an arbitrary potential satisfying the requirements listed above. In Sec.~\ref{sec:SOC} we briefly review the basics of Rashba spin-orbit coupling and establish our notation. In Sec.~\ref{sec:scatt} we formulate the Lippmann-Schwinger equation for our problem, introduce the $T$-matrix and establish its relation to the $S$-matrix, then relate the $T$ and $S$ matrices to the differential and total scattering cross sections, deriving a new optical theorem for low-energy Rashba scattering in the process. In Sec.~\ref{sec:RashbaT} we present a nonperturbative solution of the Lippmann-Schwinger equation, obtaining the $T$-matrix in the low-energy limit. Our solution relies on the application of a momentum cutoff around the degenerate low-energy Rashba ring of states. As expected, the low-energy $T$-matrix is universal and exhibits a distinct 1D character. Using the relation between the $S$ and $T$ matrices derived earlier, we obtain the universal off-diagonal $S$-matrix of Ref.~\cite{Hutchinson2016}. Our optical theorem allows us to show that the quantized plateaus seen in our previous work for the hard disk potential are indeed a generic feature of the low-energy total cross section, independent of the details of the potential. Finally, in Sec.~\ref{sec:examples} we illustrate these results with a number of example potentials. We conclude in Sec.~\ref{sec:conclusion}, and derive a number of technical results in Appendices~\ref{App:AppendixA}-\ref{app:RashbaT}.

\section{Rashba spin-orbit coupling}
\label{sec:SOC}
We begin with the unperturbed Rashba Hamiltonian in two dimensions~\cite{bychkov1984},
\be
H(\b{k})=\frac{\b{k}^2}{2m}+\lambda\hat{\b{z}}\cdot(\b{\sigma}\times\b{k}),
\ee
where $\b{k}$ is the electron wave vector confined to the $x$-$y$ plane (we work in units of $\hbar=1$), $\b{\sigma}$ is the vector of Pauli matrices, and $\lambda$ is the Rashba coupling. This Hamiltonian is readily diagonalized to give the spin-split spectrum
\be
E_{\pm}(\b{k})=\frac{\b{k}^2}{2m}\pm\lambda|\b{k}|,
\ee
and eigenspinors
\be\label{eq:eta}
\eta^\pm(\theta_\b{k})=\frac{1}{\sqrt{2}}
\begin{pmatrix}
1 \\
\mp ie^{i\theta_\b{k}}
\end{pmatrix}.
\ee
There is a degenerate ring of states for each wave vector of magnitude $k\equiv|\b{k}|$. Since the spin expectation value in the corresponding eigenstates is locked orthogonally to the wave vector, this spectrum consists of two bands of opposite \emph{helicity}, designated by the $\pm$ subscripts. We are exclusively interested in the lower of these two bands, and so it is useful to write all quantities in terms of the ground state energy $-E_0\equiv -m\lambda^2/2$, and the ground state wave vector magnitude $k_0\equiv m\lambda$. These are the only quantities that are controlled by the Rashba coupling in our problem. Along this vein, we parameterize the electron scattering energy by the dimensionless quantity $\delta\equiv\sqrt{1-|E|/E_0}$.

For any given energy $-E_0<E<0$ and wave vector angle $\theta$, there exist two degenerate negative-helicity states of different wave vector magnitude. One has a wave vector whose magnitude is greater than $k_0$, while the other is less than $k_0$. We denote these magnitudes by
\be\label{eq:k><}
k_\gtrless=k_0(1\pm\delta).
\ee


\section{Scattering Quantities}
\label{sec:scatt}

Roughly speaking, the $T$-matrix is the portion of the $S$-matrix in which some scattering occurs. Since Rashba scattering involves some subtleties, it is worth deriving the exact relation between these objects in the negative energy regime, elucidating various scattering quantities along the way. The natural starting point is the Lippmann-Schwinger equation,
\begin{eqnarray}\label{eq:Lipp}
\psi_{\b k \sigma}(\b{r};E)&=&\psi^{\rm in}_{\b k \sigma}(\b{r};E)\nonumber\\
&&+\sum_{\sigma'}\int d^2\b r' G_{\sigma\sigma'}^+(\b r, \b r'; E)V(\b r')\psi_{\b k \sigma'}(\b r'),\nonumber\\
\end{eqnarray}
where $G^+(\b r, \b r'; E)$ is the retarded position-space Green's function of the unperturbed Hamiltonian, $V(\b r)$ is the scattering potential, and $\sigma$ is a spin index. The incident wavefunction is chosen to be a negative helicity plane wave with wavevector $\b{k}$ oriented at an angle $\theta_\b{k}$ with respect to the $x$-axis,
\be
\psi^{\rm in}_{\b{k}}(\b{r};E)=e^{i\b{k}\cdot \b{r}}\eta^-(\theta_\b{k}).
\ee

We can relate this to the $T$-matrix through the defining relation
\be
T|i\rangle=V|\psi\rangle,
\ee
where $|i\rangle$ is the initial state, and $|\psi\rangle$ is the scattering state. In terms of wavefunctions, we write
\be
V|\psi\rangle=\sum_{\sigma'}\int d^2\b r' T|\b r'\sigma'\rangle\psi^{\rm in}_{\b k \sigma'}(\b{r}';E),
\ee
or equivalently,
\be\label{eq:V(r)}
V(\b r)\psi_{\b k \sigma}(\b{r};E)=\sum_{\sigma'}\int d^2\b r' T^{\b r\b r'}_{\sigma\sigma'}e^{i\b k\cdot\b r'}\eta_{\sigma'}^-(\theta_\b{k}).
\ee
We will need to Fourier transform the $T$-matrix to momentum-space,
\be\label{eq:TFourier}
T^{\b r\b r'}_{\sigma\sigma'}=\int\frac{d^2\b{k}'}{(2\pi)^2}\int\frac{d^2\tilde{\b{k}}}{(2\pi)^2}T^{\b{k}'\tilde{\b{k}}}_{\sigma\sigma'}e^{i\b k'\cdot\b r}e^{-i\tilde{\b{k}}\cdot \b r'}.
\ee
Substituting \eqref{eq:TFourier} into \eqref{eq:V(r)} and \eqref{eq:V(r)} into \eqref{eq:Lipp}, we obtain a modified Lippman-Schwinger equation,
\begin{eqnarray}
\psi_{\b k \sigma}(\b{r};E)&=&\psi^{\rm in}_{\b k \sigma}(\b{r};E)\nonumber\\
&&+\sum_{\sigma'\sigma''}\int d^2\b r'\int\frac{d^2\b k'}{(2\pi)^2} G_{\sigma\sigma'}^+(\b r, \b r'; E)\nonumber\\
&&\times T^{\b{k}'\b{k}}_{\sigma'\sigma''}e^{i\b{k}'\cdot \b{r}'}\eta^-_{\sigma''}(\theta_\b{k}).\nonumber\\
\label{eq:Lipp2}
\end{eqnarray}
To proceed any further requires knowing the position-space Green's function. This is derived in Appendix \ref{App:AppendixA}: see Eq.~\eqref{eq:GDiag} for $\sigma=\sigma'$ and Eq.~\eqref{eq:GOff} for $\sigma\neq\sigma'$. To match with the $S$-matrix, we must consider the asymptotic wavefunction, which for a finite range potential, amounts to imposing $r\gg r'$, $|\b{r}-\b{r}'|\approx r-\hat{\b{r}}\cdot\b{r}'$ and $\theta_{\b{r}-\b{r}'}\approx\theta_\b{r}$ in the Green's function, where $\hat{\b{r}}$ denotes a unit vector in the direction of $\b{r}$. Using the asymptotic form of the Hankel function for large argument,
\be
H_l^\pm(x)\approx\sqrt{\frac{2}{\pi x}}e^{\pm i(x-l\pi/2-\pi/4)},
\ee
we obtain the asymptotic Green's function
\be\label{eq:Greenposition}
G^+_{\sigma\sigma'}(\b{r},\b{r}';E)\approx-\frac{m}{k_++k_-}\sqrt{\frac{i}{2\pi r}}\sum_{j=+,-}g^j_{\sigma\sigma'}(\b{r})e^{-i\b{k}_j\cdot \b{r}'},
\ee
where $\b{k}_j\equiv k_j\hat{\b{r}}$ and $k_j$ is given in Eq.~\eqref{eq:kHel} and \eqref{eq:kHel2}. We have also defined the matrix
\begin{eqnarray}
g^j(\b{r})&\equiv&\sqrt{k_j}e^{ik_jr}
\begin{pmatrix}
1 & ie^{-i\theta_\b{r}}j \\
-ie^{i\theta_\b{r}}j & 1\\
\end{pmatrix}\nonumber\\
&=&2\sqrt{k_j}e^{ik_jr}\eta^j(\theta_\b{r})\eta^{j}(\theta_\b{r})^\dagger.
\end{eqnarray}

Since the $\b{r}'$ dependence of the Green's function has been isolated, we can now evaluate the integrals in \eqref{eq:Lipp2} to get the asymptotic wavefunction,
\begin{eqnarray}\label{eq:LippFinal}
\psi_{\b k}(\b{r};E)&\approx&\psi^{\rm in}_{\b k}(\b{r};E)-\frac{m}{(k_++k_-)}\sqrt{\frac{2i}{\pi r}}\nonumber\\
&&\times\sum_{j=+,-}\sqrt{k_j}e^{ik_jr}\eta^j(\theta_\b{r})\eta^j(\theta_\b{r})^{\dagger}T^{\b{k}_j\b{k}}\eta^-(\theta_\b{k}).\nonumber\\
\end{eqnarray}

\subsection{Relation between $T$ and $S$ matrices}
At this point we orient the $x$-axis along the incident wave direction ($\theta_\b{k}=0$) and recognize that for any negative energy, the magnitude of the corresponding wavevector is either $k_>$ or $k_<$ using the notation in \eqref{eq:k><}. We write this as $\b{k}=k_\mu\hat{\b{x}}$, where $\mu=>,<$.

To connect the $T$-matrix to the $S$-matrix (or equivalently the scattering amplitude), we use the definition of the $S$-matrix as the unitary transformation from asymptotic ingoing to asymptotic outgoing states. Schematically,
\begin{eqnarray}
\psi_>(\b{r};E)&\sim&\psi^{\rm in}_>+S_{>>}\phi_>^{\rm out}+S_{<>}\phi_<^{\rm out},\\
\psi_<(\b{r};E)&\sim&\psi^{\rm in}_<+S_{><}\phi_>^{\rm out}+S_{<<}\phi_<^{\rm out}.
\end{eqnarray}  
In Ref.~\cite{Hutchinson2016}, the form of the $S$-matrix for lower-helicity scattering off of a finite range, circularly symmetric potential was obtained.
Using a slightly modified notation, we summarize these results by writing the asymptotic wavefunction outside such a potential as
\begin{eqnarray}\label{eq:wavefunc}
\psi_\mu(\b{r};E)&\approx&\psi^{\rm in}_\mu(\b{r};E)\nonumber\\
&&+2m\sqrt{\frac{i}{k_\mu}}\sum_{\nu=>,<} f_{\mu\nu}(\theta_\b{r})\frac{e^{is_\nu k_\nu r}}{\sqrt{r}}\eta^{s_\nu}(\theta_\b{r}).\nonumber\\
\end{eqnarray}
Here, the indices $\mu, \nu$ indicate the magnitude of the wavevector $k_\gtrless$ as discussed above, $s_\mu\equiv\text{sgn}(k_\mu -k_0)$, and $\psi_\mu(\b{r};E)\equiv\psi_{\b{k}}(\b{r})|_{\b{k}=k_\mu\hat{\b{x}}}$. The common spinor factor $\eta^{s_\mu}(\theta)$ is formally equivalent to the definition \eqref{eq:eta} due to the fact that the group velocity is oppositely directed for the $<$ state (see Ref.~\cite{Hutchinson2016} for details). The factor of $2m\sqrt{i/k_\mu}$ in front of the sum is chosen to make $f_{\mu\nu}$ consistent with the conventional \emph{scattering amplitude} in two dimensions~\cite{adhikari1986}. 
With these conventions, the scattering amplitude has the following relation to the $S$-matrix expanded in partial waves,
\be\label{eq:fS}
f_{\mu\nu}(\theta_\b{r})=\frac{e^{-\frac{i\pi}{4}(1+s_\nu)}}{4m}\sqrt{\frac{2}{\pi}}\sum_{l=-\infty}^\infty e^{il(\theta_\b{r}+\frac{\pi}{2}(1-s_\nu))}(S^l_{\mu\nu}-\mathbb{I}_{\mu\nu}).
\ee

The strategy now is to simply equate \eqref{eq:LippFinal} and \eqref{eq:wavefunc}. To do this, we need a sum over wave vector magnitudes $\nu=>,<$ in \eqref{eq:LippFinal} rather than helicity index $j=+,-$. This is accomplished by noting from \eqref{eq:k><} and \eqref{eq:kHel2}, the mathematical relation
\be
k_\pm=\mp k_\lessgtr,
\ee
valid for any negative energy. Eq.~\eqref{eq:LippFinal} then reads
\begin{eqnarray}
\psi_{\mu}(\b{r};E)&\approx&\psi^{\rm in}_{\mu}(\b{r};E)-\frac{m}{k_>-k_<}\sqrt{\frac{2i}{\pi r}}\nonumber\\
&&\times\bigl(\sqrt{k_>}e^{ik_>r}\eta^-(\theta_\b{r})\eta^-(\theta_\b{r})^{\dagger}T^{\b{k}_>\b{k}}\eta^-(0)\nonumber\\
&&+i\sqrt{k_<}e^{-ik_<r}\eta^+(\theta_\b{r})\eta^+(\theta_\b{r})^{\dagger}T^{-\b{k}_<\b{k}}\eta^-(0)\bigr),\nonumber\\
\end{eqnarray}
where $\b{k}_{\gtrless}\equiv k_{\gtrless}\hat{\b{r}}$.
For the $k_>$ term, we simply note that since $\theta_{\b{k}_>}=\theta_\b{r}$ and $\theta_\b{k}=0$, 
\be
\eta^-(\theta_\b{r})^{\dagger}T^{\b{k}_>\b{k}}\eta^-(0)=T^{\b{k}_>\b{k}}_{--},
\ee
which is the component of the helicity transform of $T$ involving only transitions within the negative helicity state. For the $k_<$ term, we use the fact that $\theta_{-\b{k}_<}=\theta_{\b{r}}+\pi$ 
to write the eigenspinors as
\be
\eta^+(\theta_\b{r})=\frac{1}{\sqrt{2}}
\begin{pmatrix}
1 \\
-ie^{i\theta_\b{r}}
\end{pmatrix} =
\frac{1}{\sqrt{2}}\begin{pmatrix}
1 \\
ie^{i(\theta_\b{r}+\pi)}
\end{pmatrix}=\eta^-(\theta_{-\b{k}_<}),
\ee
which makes it clear that
\be
\eta^+(\theta_\b{r})^{\dagger}T^{-\b{k}_<\b{k}}\eta^-(0)=T^{-\b{k}_<\b{k}}_{--}.
\ee
The Lippman-Schwinger equation finally reads
\begin{eqnarray}
\psi_{\mu}(\b{r};E)&\approx&\psi^{\rm in}_{\mu}(\b{r};E)+\frac{me^{-\frac{i\pi}{4}}}{k_<-k_>}\sqrt{\frac{2i}{\pi r}}\nonumber\\
&&\times\sum_\nu\sqrt{k_\nu}e^{is_\nu(k_\nu r+1)}T_{--}^{s_\nu \b{k}_\nu \b{k}_\mu}\eta^{s_\nu}(\theta_\b{r})e^{-\frac{i\pi}{4}s_\nu}.\nonumber\\\label{eq:Lipp3}
\end{eqnarray}\
Comparing \eqref{eq:Lipp3} to \eqref{eq:wavefunc}, we may simply read off the relation between the $T$-matrix and scattering amplitude:
\be
T_{--}^{s_\nu\b{k}_\nu \b{k}_\mu}=\sqrt{2\pi}\frac{(k_<-k_>)}{\sqrt{k_\mu k_\nu}}e^{-\frac{i\pi}{4}(1+s_\nu)} f_{\mu\nu}(\theta_\b{r}),
\ee
or, in terms of the $S$-matrix written in \eqref{eq:fS},
\be\label{eq:TS}
T_{--}^{\b{k}_\nu \b{k}_\mu}=\frac{i}{m}\frac{k_0\delta}{\sqrt{k_\mu k_\nu}}\sum_{l=-\infty}^\infty e^{il\theta}(S^l_{\mu\nu}-\mathbb{I}_{\mu\nu}),
\ee
using $k_>-k_<=2k_0\delta$, and letting $\theta\equiv\theta_\b{r}=\theta_{\b{k}_\nu}-\theta_{\b{k}_\mu}$.
Rotational symmetry of the Hamiltonian allows us to expand the $T$-matrix in partial wave components as well, so that we may invert \eqref{eq:TS} to get
\be\label{eq:ST}
S^l_{\mu\nu}=\mathbb{I}_{\mu\nu}-\frac{im}{k_0\delta}\sqrt{k_\mu k_\nu}T^l(k_\nu,k_\mu),
\ee
where $T^{\b{k}_\nu \b{k}_\mu}_{--}=\sum_{l=-\infty}^\infty T^l(k_\nu,k_\mu)e^{il\theta}$.
The above result can be shown to be equivalent to the usual definition of the $S$-matrix (see, e.g., Ref.~\cite{sakurai1995}),
\be\label{eq:TS2}
S_{fi}=\delta_{fi}-2\pi i\delta(E_f-E_i)T_{fi},
\ee
with the appropriate change of basis (see Appendix \ref{App:AppendixB} for details). 
\subsection{Cross section and optical theorem}
To complete our scattering formalism we determine the differential cross section. Beginning with Fermi's golden rule, the transition rate is connected to the $T$-matrix via
\be
w_{\mu\rightarrow\nu}d\theta=2\pi|T^{\b{k}_\mu \b{k}_\nu}_{--}|^2\rho(E_\nu)d\theta,
\ee
where $\rho(E_\nu)$ is the density of final states in the channel $\nu$ within an angle $d\theta$ of $\theta$:
\be
\rho(E_\nu)=\int_0^\infty\frac{dk\,k}{(2\pi)^2}\delta(E_\nu-E(k))=\frac{m}{(2\pi)^2}\frac{k_\nu}{k_0\delta}.
\ee
Furthermore, the differential cross section in this channel is simply the transition rate divided by the incident flux,
\begin{eqnarray}
\frac{d\sigma}{d\theta}\bigg|_{\mu\nu}&=&\frac{w_{\mu\rightarrow\nu}}{|\b{j}_\mu|}\nonumber\\
&=&\frac{m^2}{2\pi}\frac{k_\nu}{k_0^2\delta^2}|T^{\b{k}_\mu \b{k}_\nu}_{--}|^2\nonumber\\
&=&\frac{1}{2\pi k_\mu}\left|\sum_{l=-\infty}^\infty e^{il\theta}(S^l_{\mu\nu}-\mathbb{I}_{\mu\nu})\right|^2.
\end{eqnarray}
This last expression was denoted $|\phi_{\mu\nu}|^2$ in Ref.~\cite{Hutchinson2016}. Integrating over angles and summing over scattering channels gives the total cross section for an incident $k_\mu$ wave,
\begin{eqnarray}
\sigma_\mu&=&\int_0^{2\pi}\sum_\nu \frac{1}{2\pi k_\mu}\left|\sum_{l=-\infty}^\infty e^{il\theta}(S^l_{\mu\nu}-\delta_{\mu\nu})\right|^2\\
&=&\frac{1}{k_\mu}\sum_{l=-\infty}^\infty \left(\left|S^l_{\mu\mu}-1\right|^2+\left|S^l_{\mu,-\mu}\right|^2\right)\\
&=&\frac{1}{k_\mu}\sum_{l=-\infty}^\infty (2-(S^l_{\mu\mu}+S_{\mu\mu}^{l*}))\label{eq:unitarity}\\
&=&\frac{2}{k_\mu}\sum_{l=-\infty}^\infty (1-\re(S^l_{\mu\mu})),
\end{eqnarray}
where $S^l_{\mu,-\mu}$ denotes the off-diagonal component with first index $\mu$, and we used the unitarity condition of the $S$-matrix ($|S^l_{\mu,-\mu}|^2=1-|S_{\mu\mu}|^2$) in line \eqref{eq:unitarity}. The final form of this cross section makes it clear that the diagonal part of the $T$-matrix in \eqref{eq:TS} obeys an optical theorem, since
\begin{eqnarray}
\im T^{\b{k}_\mu \b{k}_\mu}_{--}(\theta=0)&=&-\frac{k_0\delta}{mk_\mu}\sum_{l=-\infty}^\infty (1-\re(S^l_{\mu\mu}))\nonumber\\
&=&-\frac{k_0\delta}{2m}\sigma_\mu\label{eq:cross}.
\end{eqnarray}

\section{Rashba $T$-matrix}\label{sec:RashbaT}
With this scattering formalism at hand, we may compute any scattering observable in a Rashba system with $E<0$, provided we know the  $T$-matrix $T^{\b{k}_\mu \b{k}_\nu}_{--}$. In a conventional 2D system without spin-orbit coupling, the $T$-matrix takes on a form at low energies that is dominated by the s-wave term,
\be\label{eq:convent}
T^{\b{kk}'}\approx T^0(E)\sim\frac{1/m}{i-\frac{1}{\pi}\ln(E/E_a)},
\ee 
where $E_a$ is a parameter that encodes the potential $V$, and is related to the scattering length (see, e.g., Ref.~\cite{friedrich2013,randeria1990}). Before doing any calculation, we can already see that the Rashba $T$-matrix must have a different energy dependence than (\ref{eq:convent}), simply by looking at the Lippmann-Schwinger equation \eqref{eq:Lipp3}. Since the coefficient of the scattered wavefunction goes as $1/\delta$ for low energies, the $T$-matrix must at least be linear in $\delta$ in order to keep the probability density finite. We now make this explicit by deriving the low-energy Rashba $T$-matrix for any circularly symmetric, spin-independent potential of finite range.

First, we impose a momentum cutoff
\be\label{eq:cutoff}
k_0-\tilde\Lambda<k<k_0+\tilde\Lambda,
\ee
to avoid ultraviolet divergences. This amounts to keeping only the low-energy modes in our model, similar to the momentum shell renormalization group approach in the many-body problem~\cite{shankar1994,yang2006}. The appropriate dimensionless quantity corresponding to this cutoff is $\Lambda\equiv\tilde{\Lambda}/k_0$, so that we will always enforce the following hierarchy of scales:
\be
\delta\ll\Lambda\ll1.
\ee
   \begin{figure}[t]
	\centering
	\includegraphics[width=\columnwidth]{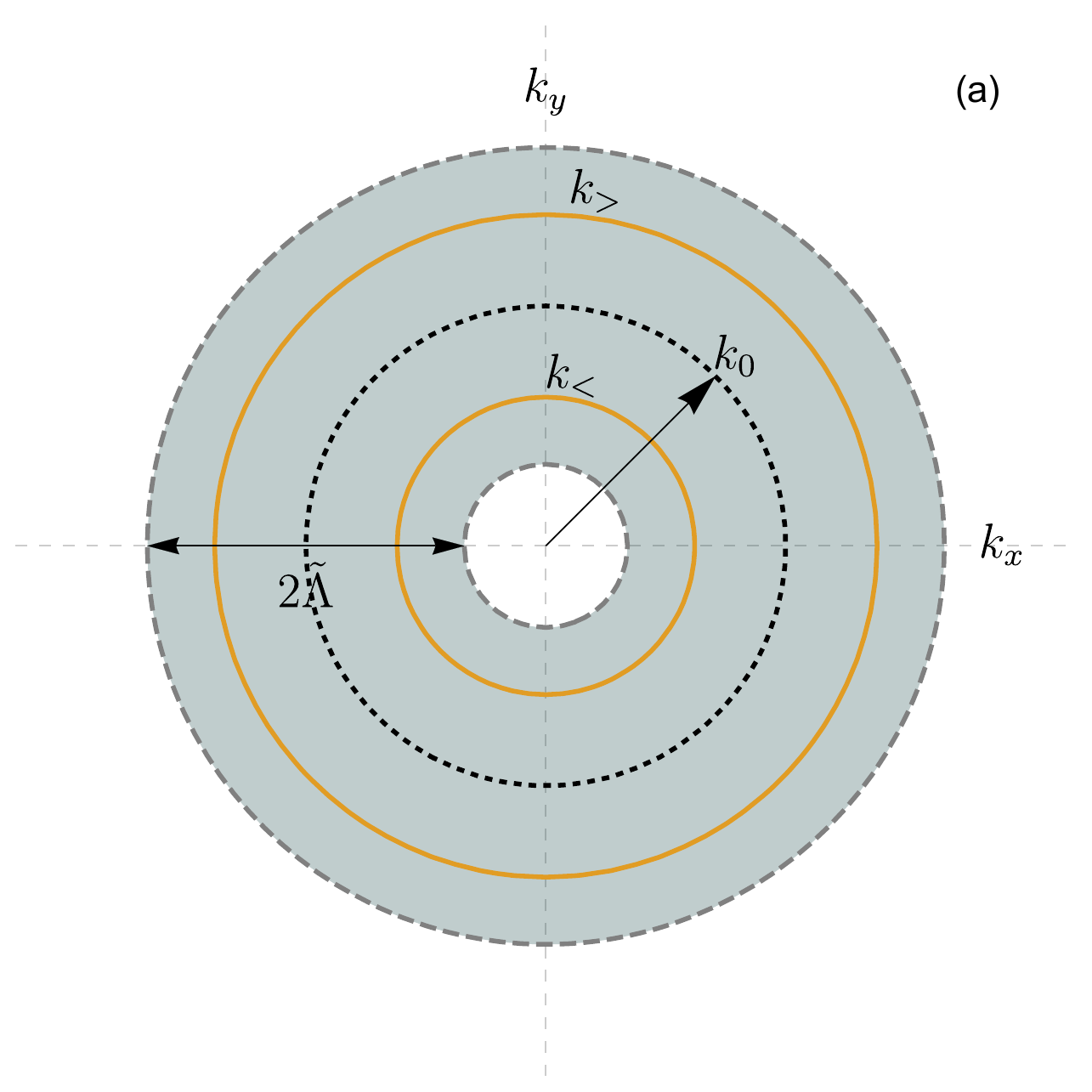} 
	\includegraphics[width=\columnwidth]{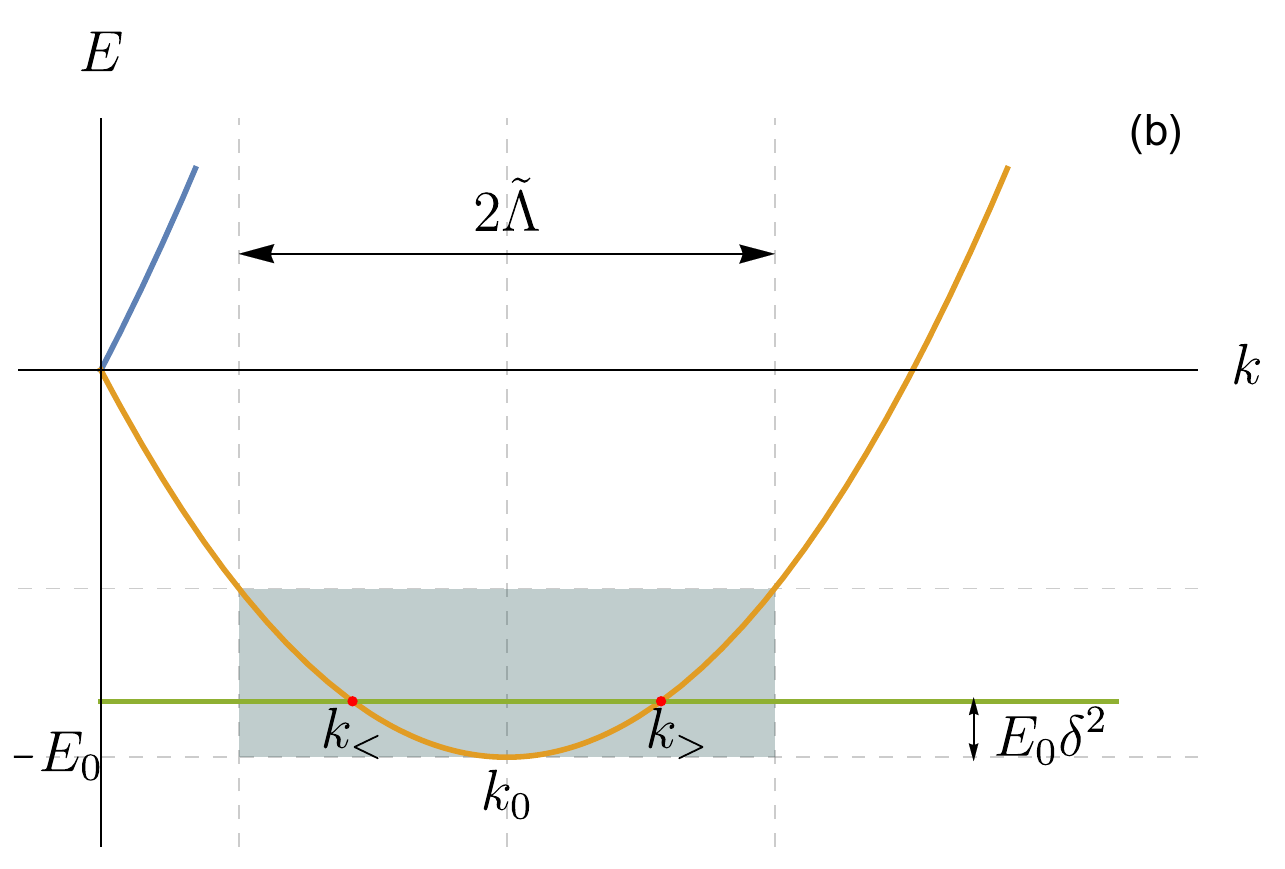} 
\caption{(a) Constant energy contours in momentum space and (b) low-energy spectrum for a single Rashba electron. The shaded region shows the allowed virtual transitions with $|k-k_0|<\tilde\Lambda$ to be incorporated in the $T$-matrix. The orange lines show the continuum of negative helicity eigenstates. The blue line in (b) is the positive helicity branch.}
\label{fig:kspace}
\end{figure}

In the helicity basis denoted by $i,j$, any central spin-independent potential may be written as
\begin{eqnarray}
V_{ij}(\b{k,k}')&=&\int d^2\b{r}\,e^{i(\b{k}-\b{k}')\cdot\b{r}}V(\b{r})\eta^i(\theta_{\b{k}'})^\dagger\eta^j(\theta_\b{k})\nonumber\\
&=&\frac{1}{2}\sum_{l=-\infty}^\infty V^l(k,k')e^{il(\theta_{\b{k}'}-\theta_\b{k})}\left(1+ije^{i(\theta_\b{k}-\theta_{\b{k}'})}\right)\nonumber\\
&=&\frac{1}{2}\sum_{l=-\infty}^\infty\left(V^l(k,k')+ijV^{l+1}(k,k')\right)e^{il\theta_{\b{k}'-\b{k}}},\nonumber\\
\end{eqnarray}
where $\theta_{\b{k}'-\b{k}}\equiv\theta_{\b{k}'}-\theta_\b{k}$, and in the second line, we introduced the partial wave component
\be\label{eq:Vl}
V^l(k,k')
=\int^{2\pi}_0\frac{d\theta_{\b{k}'-\b{k}}}{2\pi}\int^\infty_0 dr\,rV(r)J_0(|\b{k}-\b{k}'|r)e^{il\theta_{\b{k}'-\b{k}}},
\ee
where $J_0(|\b{k}-\b{k}'|r)$ is the zeroth order Bessel function of the first kind.

Now the $T$-matrix is defined by the Born series
\be
T=V+VG^+T.
\ee
We write this in the momentum-helicity basis $|\b{k},i\rangle$ in which the Green's function is diagonal,
\begin{eqnarray}\label{eq:TBorn}
T_{ji}^{\b{k}_\nu \b{k}_\mu}&=&V_{ji}(\b{k}_\nu,\b{k}_\mu)\nonumber\\
&&+\sum_{n=+,-}\int\frac{d^2q}{(2\pi)^2}V_{jn}(\b{k}_\nu,\b{q})G^+_{nn}(q)T_{ni}^{\b{q k}_\mu}.
\end{eqnarray}
We want to expand the potential about the ground state wavevector $k_0$. More precisely, let us examine the $V^l$ components given by \eqref{eq:Vl}. For the on-shell terms $V_{ji}(\b{k}_\mu,\b{k}_\nu)$ in \eqref{eq:TBorn}, the argument of this Bessel function is
\begin{eqnarray}
|\b{k}_\mu-\b{k}_\nu|r&=&r\sqrt{k_\mu^2+k_\nu^2-2k_\mu k_\nu\cos\theta_{\b{k}'-\b{k}}}\nonumber\\
&=&\sqrt{2}k_0r\sqrt{1-\cos\theta_{\b{k}'-\b{k}}}+\mathcal{O}(\delta).
\end{eqnarray}
The off-shell components in the integral of \eqref{eq:TBorn} may also be expanded about $\delta=0$. The argument of the Bessel function becomes
\be
|\b{k}_\nu - \b{q}|r=\sqrt{2}k_0r\sqrt{(1+\epsilon)(1-\cos\theta_{\b{k}'-\b{k}})}+\mathcal{O}(\delta),
\ee
where we have changed the integration variable using $q\equiv k_0(1+\epsilon)$. Thus, to order $\delta$ in the potential, we can approximate the on-shell terms as $V_{ji}(\b{k}_\nu,\b{k}_\mu)\approx V_{ji}(k_0\hat{\b{k}}_\nu,k_0\hat{\b{k}}_\mu)$, and the off-shell terms as $V_{ji}(\b{k}_\nu,\b{q})\approx V_{ji}(k_0\hat{\b{k}}_\nu,\b{q})$. This is a crucial approximation. Since now the right hand side of \eqref{eq:TBorn} is independent of the magnitude $k_\nu$, the $T$-matrix is independent of this magnitude as well
\be
T_{ij}^{\b{k}_\nu \b{k}_\mu}\approx T_{ij}(\hat{\b{k}}_\nu,\b{k}_\mu).
\ee
We will argue in Appendix \ref{app:RashbaT} that the error in this approximation is $\mathcal{O}(\delta^2)$.
Writing the $T$-matrix in partial wave components just as we did with the potential, the Born series simplifies to
\begin{eqnarray}\label{eq:TBornl}
\sum_{l=-\infty}^\infty T^l_{ji}(k_\mu)e^{il\theta}&=&\sum_{l=-\infty}^\infty \frac{1}{2}[V^l(k_0,k_0)+ijV^{l+1}(k_0,k_0)]e^{il\theta}\nonumber\\
&&+\sum_{n=+,-}\sum_{l=-\infty}^\infty\int_0^\infty\frac{dq\,q}{4\pi}\Bigl(V^l(k_0,q)\nonumber\\
&&+jnV^{l+1}(k_0,q)\Bigr)G^+_{nn}(q)T^l_{ni}(k_\mu)e^{il\theta}.\nonumber\\
\end{eqnarray}
Equation \eqref{eq:TBornl} may be solved algebraically for each partial wave component. Since the diagonal parts of the potential are equal, this equation decouples into two pairs of coupled equations. For the lower helicity band, the relevant pair is
\begin{eqnarray}\label{eq:T--}
T^l_{--}(k_\mu)&\approx&\frac{1}{2}[V^l(k_0,k_0)+V^{l+1}(k_0,k_0)]\nonumber\\
&&+I^l_-T^l_{--}(k_\mu)+J^l_-T^l_{+-}(k_\mu),\\
T^l_{+-}(k_\mu)&\approx&\frac{1}{2}[V^l(k_0,k_0)-V^{l+1}(k_0,k_0)]\nonumber\\
&&+I^l_+T^l_{--}(k_\mu)+J^l_+T^l_{+-}(k_\mu),
\end{eqnarray}
where we have defined the integrals
\begin{eqnarray}
I^l_\pm&=&\int_0^\infty\frac{dq\,q}{4\pi}[V^l(k_0,q)\mp V^{l+1}(k_0,q)]G^+_{--}(q),\label{eq:Il}\\
J^l_\pm&=&\int_0^\infty\frac{dq\,q}{4\pi}[V^l(k_0,q)\pm V^{l+1}(k_0,q)]G^+_{++}(q).
\end{eqnarray}
Using the fact that $J_-I_+=-J_+I_-$, we may solve for $T^l_{--}$ to get
\begin{eqnarray}
T^l_{--}&\approx&\frac{1}{2(1-I^l_-(1-2J^l_+) -J^l_+)}\nonumber\\
&&\times[V^l(k_0,k_0)(1-J^l_++J^l_-)\nonumber\\
&&+V^{l+1}(k_0,k_0)(1-J^l_+-J^l_-)].
\end{eqnarray}
The $J_\pm$ integrals correspond to transitions between different helicity bands, and these are expected to have a negligible contribution to the low energy scattering. Indeed, one can show that $J_\pm\sim\Lambda\ll1$ and so
\be
T^l_{--}\approx\frac{\frac{1}{2}[V^l(k_0,k_0)+V^{l+1}(k_0,k_0)]}{1-I^l_-}.
\ee
The energy dependence of the $T$-matrix is entirely determined by the integral $I^l_-$ of Eq.~\eqref{eq:Il}. We claim that to leading order in $\delta$, this integral is approximated by
\begin{eqnarray}\label{eq:Il2}
I^l_-&=&-\frac{m}{2}\bigg(\frac{i}{\delta}+\frac{2}{\pi\Lambda}\bigg)[V^l(k_0,k_0)+V^{l+1}(k_0,k_0)]\nonumber\\
&&+\mathcal{O}(\delta)+\mathcal{O}(\Lambda),
\end{eqnarray}
so that the $T$-matrix is
\begin{eqnarray}\label{eq:Tfinal}
T^l_{--}&=&\frac{\frac{1}{2}[V^l(k_0,k_0)+V^{l+1}(k_0,k_0)]}{1+\frac{m}{2}(\frac{i}{\delta}+\frac{2}{\pi\Lambda})[V^l(k_0,k_0)+V^{l+1}(k_0,k_0)]}+\mathcal{O}(\delta^2).\nonumber\\
\end{eqnarray}
The detailed derivation of this result is left for Appendix~\ref{app:RashbaT}.
It is convenient to define a new dimensionless parameter
\be\label{eq:dstar}
\delta^*_l\equiv\frac{m}{2}\left(V^l(k_0,k_0)+V^{l+1}(k_0,k_0)\right),
\ee
such that to leading order in $\delta$, we can write the $T$-matrix as
\begin{eqnarray}
T^l_{--}&\approx&\frac{1}{m}\frac{\delta^*_l}{1+i\delta^*_l/\delta}\label{eq:TfirstOrder}\\
&=&-\frac{i\delta}{m}+\mathcal{O}(\delta^2)\label{eq:TlowE}.
\end{eqnarray}
The low-energy limit of the $S$-matrix \eqref{eq:ST} is thus simply
\be\label{universalSmatrix}
S^l=\begin{pmatrix}
0 & -1\\
-1 & 0
\end{pmatrix},
\ee
as was found in Ref.~\cite{Hutchinson2016} for an infinite circular barrier. Equation~(\ref{universalSmatrix}) is the main result of this work. It establishes that the low-energy $S$-matrix for circularly symmetric potentials in Rashba systems is completely universal, as conjectured in our earlier work: it is independent of any details of the potential, provided the latter has finite range. Thus all the conclusions drawn in Ref.~\cite{Hutchinson2016} from the particular form (\ref{universalSmatrix}) of the $S$-matrix, such as the extreme anisotropy of the differential cross section, are equally universal.

We now draw our attention to the peculiar energy dependence of the $T$-matrix in Eq.~(\ref{eq:TlowE}). Firstly, the $T$-matrix scales as the square root of the difference between the scattering energy and the ground state energy, in contrast with the inverse logarithm dependence found in conventional 2D systems \eqref{eq:convent}. Furthermore, it does not depend on the details of the potential (its range or strength), as already mentioned. 
Lastly, the partial wave components of the low-energy $T$-matrix are independent of partial wave number $l$. The usual intuition of low-energy physics being dominated by $s$-wave scattering does not apply to the Rashba system.

The energy dependence in \eqref{eq:TlowE} is very telling. Suppose we were to look for the universal form of a low-energy $T$-matrix in a 1D scattering problem with a conventional quadratic dispersion. We could follow the same reasoning used above. A finite-range on-shell potential in momentum space can be approximated by a constant at low energy,
\be
V(k,k')\approx \lim_{k,k'\rightarrow0} V(k,k')= \int_{-\infty}^\infty dx\,V(x)\equiv V.
\ee
The momentum-space $T$-matrix must again be independent of $k'$ in this approximation, so that
\begin{eqnarray}
T^{k'k}&\approx&T(k)\nonumber\\
&=&V+\left(\int_{-\infty}^\infty dx\int^\infty_{-\infty}\frac{dq}{2\pi}\frac{e^{iqx}V(x)}{E-\frac{q^2}{2m}+i\eta}\right)T(k)\nonumber\\
&=&V+\left(\frac{mi}{\sqrt{2mE}}\int^\infty_{-\infty}dx\,V(x)e^{i\sqrt{2mE}x}\right)T(k).\nonumber\\
\end{eqnarray}
If we only consider the lowest order terms in $E$, and make use of the fact that the potential is short-ranged, we get the following approximate solution for $T$,
\begin{eqnarray}
T&\approx&\frac{V}{1-\frac{im}{\sqrt{2mE}}\int^\infty_{-\infty}dx\,V(x)e^{i\sqrt{2mE}x}}\nonumber\\
&=&\frac{i}{m}\sqrt{2mE}+\mathcal{O}(E)\label{eq:T1D}.
\end{eqnarray}
Thus, provided we identify the 1D $\delta$ parameter with $\sqrt{2mE}$, we get the same $T$-matrix as in the low-energy 2D Rashba case\footnote{Unlike the 2D Rashba case, this 1D $\delta$ parameter is not dimensionless. This is simply because in two dimensions the momentum-space $T$-matrix must have units of inverse energy, while in one dimension, it is dimensionless.}. This connection suggests once again that low-energy Rashba scattering has a fundamental 1D character, independent of the details of the potential. Indeed, it was shown in Ref.~\cite{Hutchinson2016} that Eq.~(\ref{universalSmatrix}) implies only forward and backward scattering are allowed at very low energies. In other words, the wavefunction behaves like that of a particle scattering in a 1D system.

One might notice that \eqref{eq:T1D} and \eqref{eq:TlowE} differ by a minus sign. For the Rashba case, this sign ensures that the scattering cross section is positive in the optical theorem \eqref{eq:cross}. More importantly, it has interesting implications for the $S$-matrix. Looking at \eqref{eq:ST} we see that this sign guarantees that the diagonal part of the $S$-matrix vanishes as $\delta$ approaches zero.

The form of the low-energy $T$-matrix has interesting consequences for the cross section. First note that the total cross section becomes infinite at the threshold energy $-E_0$. This result is typical of 2D scattering, though the reasons for it are not. Using the optical theorem \eqref{eq:cross}, our $T$-matrix approximation gives a low-energy cross section of
\be\label{eq:sigmaapprox}
\sigma\approx\frac{2}{k_0}\sum_{l=-\infty}^\infty\frac{\delta^{*2}_l/\delta^2}{1+\delta^{*2}_l/\delta^2}.
\ee
Qualitatively speaking, there is a threshold parameter $\delta^*_l$ for each partial wave $l$. As we lower the energy, and thereby $\delta$, we pass through these points one by one. Each time the condition $\delta\lesssim\delta^*_l$ is satisfied an additional two partial waves (one for $l$ and one for $-l$) contribute to the scattering, and the cross section increases by $4/k_0$, tending to infinity stepwise as $\delta\rightarrow0$. This is unlike the conventional 2D case in which the prefactor $1/k$ blows up while the partial wave sum remains finite. Thus there generically is a series of jumps and plateaus in the cross section as a function of $\delta$ (see, e.g., Fig.~\ref{fig:xsec}). However, because $\delta^*_l$ decays as $l$ increases, these plateaus become smaller and smaller as we approach the ground state energy. The precise location of the jumps $\delta_l^*$ depends on the details of the potential via Eq.~(\ref{eq:dstar}), but the magnitude $4n/k_0$, $n=1,2,3,\ldots$ of the plateaus in the cross section is universal.
\section{Example Potentials}
\label{sec:examples}

\subsection{Delta function potential}
The simplest finite-range potential we can consider is the delta function
\be
V(r)=\frac{V_0}{r}\delta(r)\delta(\theta),
\ee
which has partial wave components $V^l(k,k')=V_0\delta_{l,0}$, from \eqref{eq:Vl}. Since this is independent of the momenta $k$ and $k'$, the T matrix is as well, and there is no need for an approximation at this level. Instead, the $T$-matrix exactly satisfies the equations
\begin{eqnarray}
T_{--}^0=\frac{V_0/2}{1-(I^0+J^0)}=T_{+-},
\end{eqnarray}
where we have made use of the fact that $I^l_+=I^l_-\equiv I^l$, and $J^l_+=J^l_-\equiv J^l$ for the delta potential. The integral $J^0$ may be ignored since (using $q=k_0(1+\epsilon)$)
\be
J^0=2mV_0\int_{-\Lambda}^\Lambda\frac{d\epsilon}{4\pi}\frac{1+\epsilon}{\delta^2-4(\epsilon+1)-\epsilon^2}\sim \mathcal{O}(\Lambda).
\ee 
The other integral evaluates to
\begin{eqnarray}
I^0&=&2mV_0\int_{-\Lambda}^\Lambda\frac{d\epsilon}{4\pi}\frac{1+\epsilon}{\delta^2-\epsilon^2+i\eta}\nonumber\\
&\approx&\frac{mV_0}{2\pi}\bigg(-\frac{i\pi}{\delta}+\frac{2}{\Lambda}\bigg),
\end{eqnarray}
so that
\be
T^0_{--}=\frac{V_0/2}{1+\frac{m}{2}(\frac{i}{\delta}+\frac{2}{\pi\Lambda})V_0}\approx\frac{1/m}{\frac{i}{\delta}+\frac{2}{\pi\Lambda}},
\ee
in agreement with \eqref{eq:Tfinal}. 
We emphasize that the lowest order contributions in $\delta$ are independent of the cutoff scale. This is in stark contrast with the conventional 2D case where the contact $T$-matrix satisfies
\begin{eqnarray}
T^0&=&V_0-V_0\int^\Lambda_0\frac{dk\,k}{\frac{k^2}{2m}-E-i\epsilon}T^0\nonumber\\
&=&V_0-\left(mV_0\ln\bigg|\frac{\Lambda^2}{2mE}\bigg|+i\pi mV_0\right)T^0\nonumber\\
&\approx&\frac{1/m}{i-\frac{1}{\pi}\ln|\frac{2mE}{\Lambda^2}|}.
\end{eqnarray}
One can understand this difference on dimensional grounds. Since $V_0$ is dimensionless, the energy must be compared to the only other scale around. In the conventional 2D case, this is the cutoff scale, which has the physical interpretation of an effective range of the potential (proportional to the scattering length). The process of acquiring this extra scale from what started as a scale-invariant problem is known as dimensional transmutation~\cite{camblong2001}. In the Rashba system this problem does not exist, since there is always an inherent scale to compare with, set by the spin-orbit coupling.   

Note from the optical theorem \eqref{eq:cross}, that the low-energy cross section for the delta function potential is finite; $\sigma_\mu=2/k_0$. This is highly atypical of 2D scattering both with and without Rashba spin-orbit coupling, where the threshold cross section is generally divergent. The fact that only $l=0$ contributes to the $T$-matrix for the contact potential is an artifact of the singular nature this potential. Next we will investigate more typical examples where all partial wave components become important at low energies.

\subsection{Circular barrier potential}
\begin{figure}[t]
	\centering
	\includegraphics[width=\columnwidth]{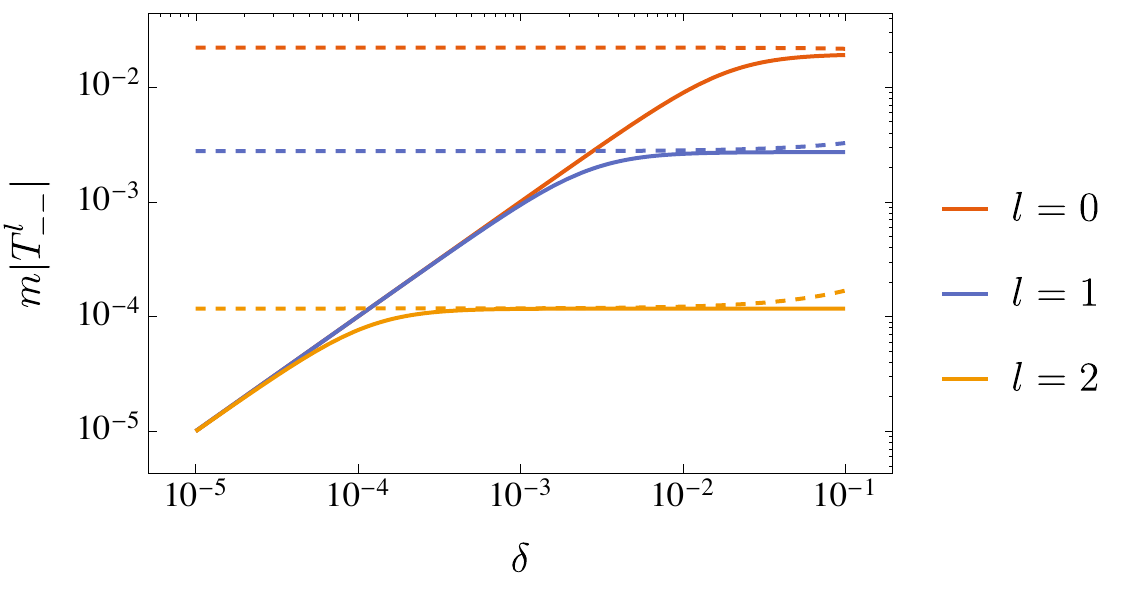} 
\caption{Absolute value of the lower-helicity $T$-matrix for the circular barrier as a function of the dimensionless parameter $\delta$, for $l=0,1,2$, obtained from \eqref{eq:Tfinal} (solid), and from the first Born approximation (dashed). The dimensionless parameters used are $mV_0R^2=0.1$, $k_0R=1$, $\Lambda=0.1$. Note that in the first Born approximation, $T^l_{--}$ is a $2\times2$ matrix in the $k_{\lessgtr}$ basis. However, these four different components are visually indistinguishable at these energies, so here we just show one of them.}
\label{fig:barrier}
\end{figure}

Consider the finite circular barrier
\be
V(r)=\begin{cases}
V_0, & r<R,\\
0, & r>R.
\end{cases}
\ee
The partial wave components 
\be
V^l(k,k')=V_0R\int_0^{2\pi}\frac{d\theta_{\b{k}'-\b{k}}}{2\pi}\frac{e^{il\theta_{\b{k}'-\b{k}}}}{|\b{k}-\b{k}'|}J_1(R|\b{k}-\b{k}'|),
\ee
are evaluated numerically. When $k=k'=k_0$, this is most easily done by summing the first few terms of \eqref{eq:f0kl}. Inserting these components into \eqref{eq:Tfinal} gives the low energy $T$-matrix which is plotted in Fig.~\ref{fig:barrier} for a short barrier. Along with our approximation, we plot the results for the first Born approximation $T^l_{--}=(V^l(k,k')+V^{l+1}(k,k'))/2$.

We see that for each $l$ component, there is a threshold energy below which the Born approximation fails to capture the correct energy dependence. The reason is most quickly seen from the asymptotic Green's function in position-space \eqref{eq:Greenposition}, which is singular at $\delta=0$ (recall that $k_++k_-=k_>-k_<=2\delta$). Evidently, it is not enough to require that the potential be perturbatively small to use the Born approximation. Instead we require $mV_0R^2/\delta<1$.

We will see below that this qualitative structure of the $T$-matrix is reproduced in the delta-shell potential, for which an exact solution is available. 

\subsection{Delta-shell potential}
\begin{figure}[t]
	\centering
	\includegraphics[width=\columnwidth]{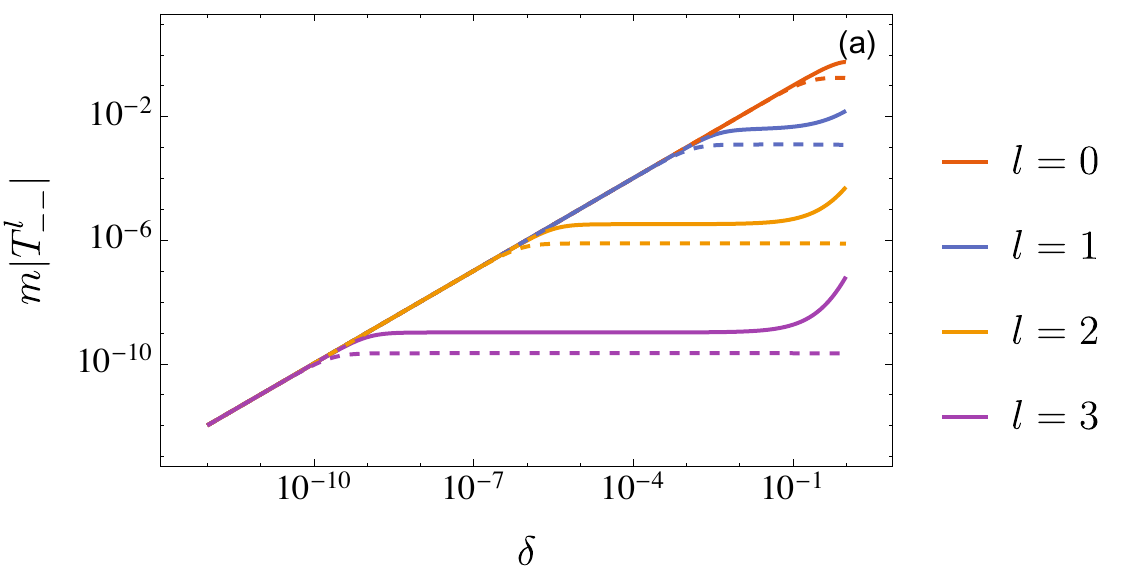} 
	\includegraphics[width=0.95\columnwidth]{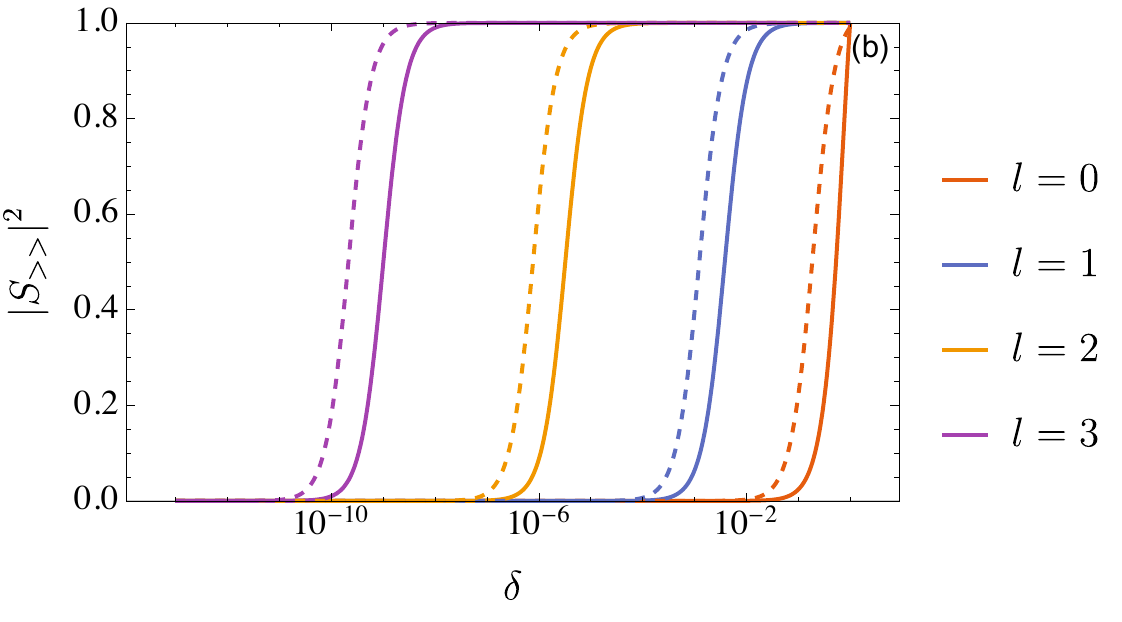} 
\caption{Absolute value of the lower-helicity $T$-matrix (a) and diagonal part of the $S$-matrix (b) for the delta-shell potential as a function of the dimensionless parameter $\delta$ for $l=0,1,2,3$. Curves are obtained from an exact calculation of the wavefunction (solid), and from the approximation \eqref{eq:Tfinal} (dashed).
The dimensionless parameters used are $mV_0R=1$, $k_0R=0.1$, $\Lambda=0.1$.}
\label{fig:shell1}
\end{figure}

We now consider the potential
\be
V(r)=V_0\delta(r-R),
\ee
so that 
\be
V^l(k,k')=V_0R\int^{2\pi}_0\frac{d\theta_{\b{k}'-\b{k}}}{2\pi}e^{il\theta_{\b{k}'-\b{k}}}J_0(|\b{k}-\b{k}'|).
\ee
Again, we plot the corresponding value of the $T$-matrix approximation (see Fig.~\ref{fig:shell1}). With this potential, we are afforded an independent check of our approximation. The $S$-matrix for the delta-shell potential was computed directly from matching conditions of the wavefunction in Ref.~\cite{Hutchinson2016}. With the aid of \eqref{eq:TS} we may translate this into the corresponding $T$-matrix (or vice versa using \eqref{eq:ST}) and compare with our approximation.

 \begin{figure}[t]
	\centering
	\includegraphics[width=\columnwidth]{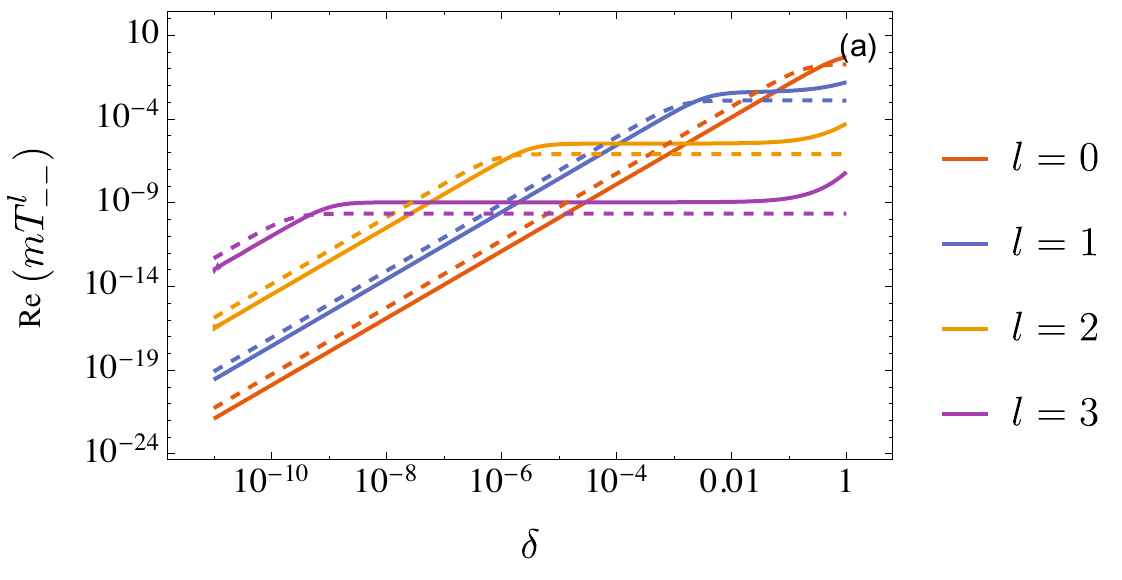} 
	\includegraphics[width=\columnwidth]{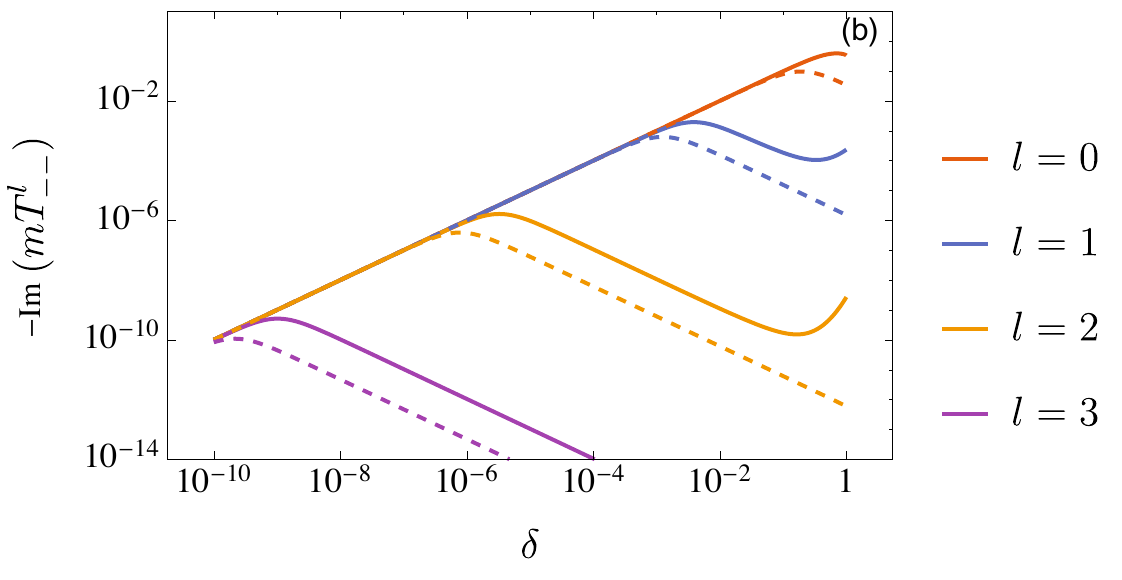} 
\caption{Real (a) and imaginary (b) parts of the lower-helicity $T$-matrix for the delta-shell as a function of the dimensionless parameter $\delta$ for $l=0,1,2,3$. Curves are obtained from an exact calculation of the wavefunction (solid), and from the approximation \eqref{eq:Tfinal} (dashed). The dimensionless parameters used are the same as in Fig.~\ref{fig:shell1}.}
\label{fig:shell2}
\end{figure}
\begin{figure}[t]
	\centering
	\includegraphics[width=\columnwidth]{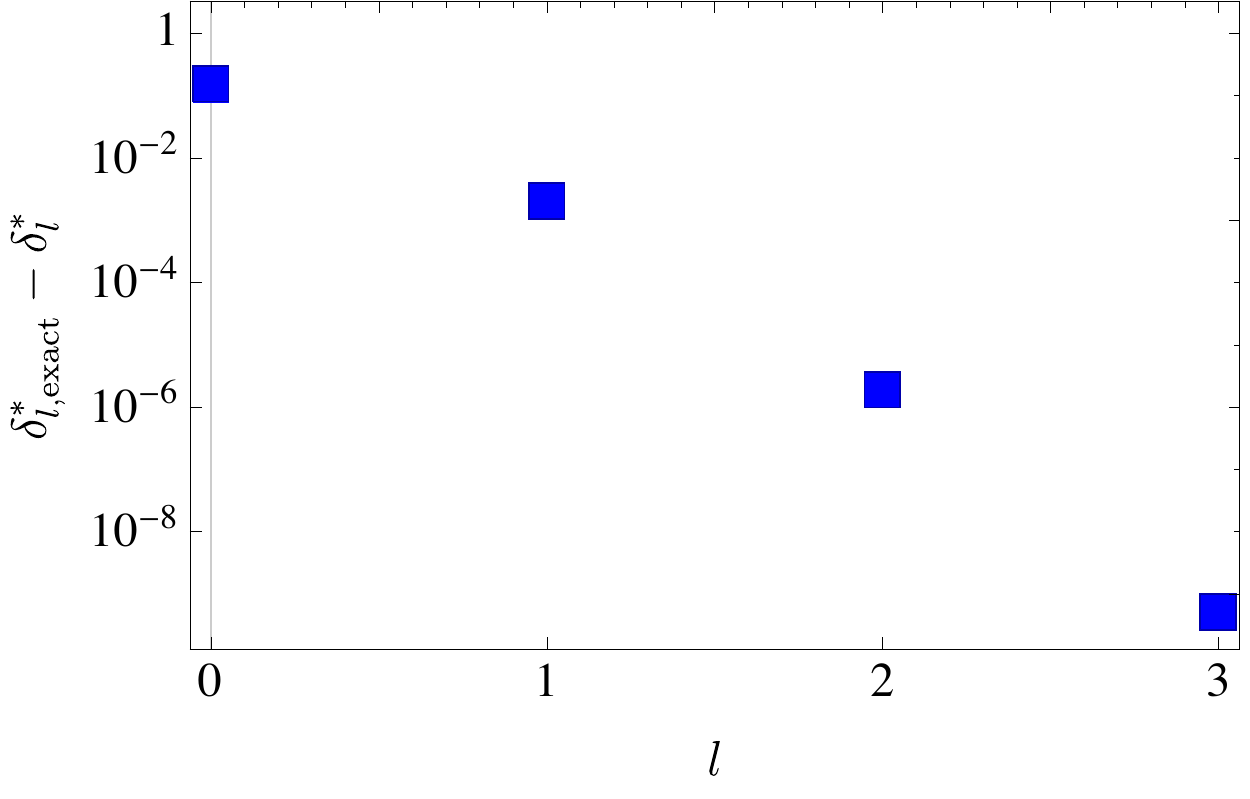} 
\caption{Difference between the value of $\delta^*_l$ computed from the exact solution of the delta-shell $T$-matrix and the value computed from our approximation, for each partial wave. The dimensionless parameters used are the same as in Fig.~\ref{fig:shell1}~and~\ref{fig:shell2}.}
\label{fig:maxdiff}
\end{figure}

Figure \ref{fig:shell2} shows the real and imaginary parts of the $T$-matrix. Note that the apparent steps in the imaginary part of the $T$-matrix translate into quantized steps in the total cross-section upon adding partial waves and using \eqref{eq:cross}. The maxima of $-{\rm Im}(mT^l_{--})$ provide a useful measure of the value of $\delta$ (for each $l$) below which the first Born approximation fails. From our approximation \eqref{eq:TfirstOrder}, this value turns out to be simply $\delta^*_l$ defined in \eqref{eq:dstar}.
In Fig.~\ref{fig:maxdiff}, we compare this to the exact value determined from the solution of $\frac{d}{d\delta}{\rm Im}(mT^l_{--})=0$, computed numerically from the exact $T$-matrix. Note that the $y$-axis in Fig.~\ref{fig:maxdiff} has a logarithmic scale: the absolute accuracy of our approximation (\ref{eq:TfirstOrder}) increases exponentially with partial wave number. However, it should be noted that the relative accuracy $(\delta^*_l-\delta_{l, {\rm exact}})/\delta_{l, {\rm exact}}$ saturates at fixed value as $l$ increases.

\begin{figure}[t]
	\centering
	\includegraphics[width=\columnwidth]{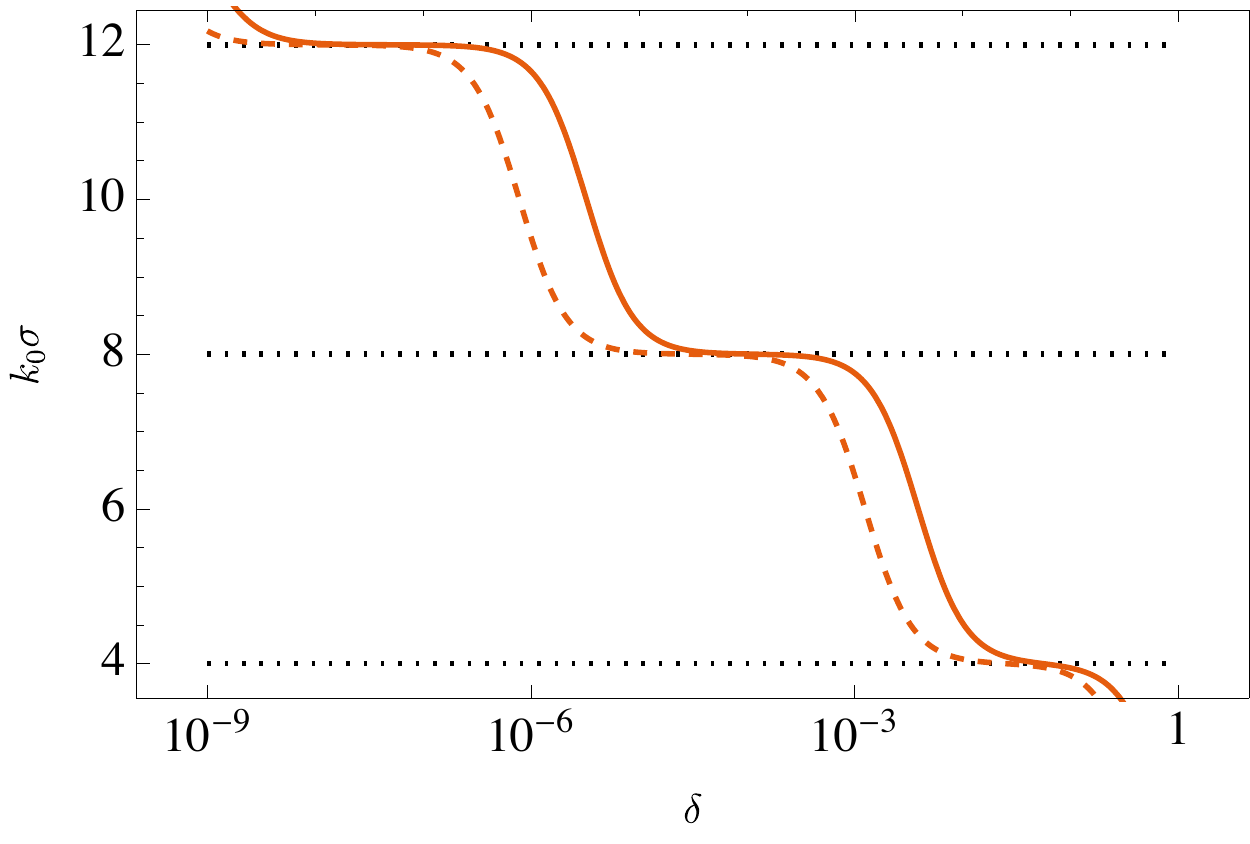} 
\caption{Total cross section $\sigma$ as a function of $\delta$ for the delta-shell potential. Solid curves show the exact result computed from the optical theorem. Dashed curves are obtained from the approximation \eqref{eq:sigmaapprox}. The dotted black lines are guides to the eye, showing that the cross section increases in steps of $4/k_0$. The dimensionless parameters used are the same as in Fig.~\ref{fig:shell1}-\ref{fig:maxdiff}.}
\label{fig:xsec}
\end{figure}

Finally, the delta-shell cross section is shown in Fig.~\ref{fig:xsec}. We see the generic features discussed in Sec.~\ref{sec:RashbaT}, namely the jumps in the cross section corresponding to the points $\delta=\delta^*_l$. Note that the $x$-axis has a logarithmic scale, so the cross section is indeed divergent at $\delta=0$. As already mentioned in the previous paragraph, this also implies the accuracy of the approximation (\ref{eq:TfirstOrder}) for the $T$-matrix increases exponentially as the energy is lowered towards the band bottom $\delta=0$.

\section{Conclusion}
\label{sec:conclusion}

We have shown that in a 2D system with Rashba spin-orbit coupling, the low-energy $T$-matrix of a particle scattering off of a circular, finite-range, spin-independent potential takes on a universal form given by \eqref{eq:TlowE}. From this, a universal form of the $S$-matrix was extracted using a complete scattering formalism developed for Rashba scattering at negative energies. These results have several important features. The $T$-matrix has a square root dependence on the difference between the energy of the scattering particle and the ground state energy, with a subleading dependence on the details of the potential. This is unlike the conventional inverse logarithm energy dependence seen in regular 2D systems, but agrees with the energy dependence of a 1D system. Indeed, this feature cannot be recovered via the Born approximation even with a perturbatively small potential, but requires a nonperturbative solution of the Lippmann-Schwinger equation. For each partial wave there exists a threshold energy below which the corresponding component of the $T$-matrix takes on this nontrivial square-root dependence on the energy. By deriving an optical theorem for negative-energy Rashba systems, we found that at these discrete threshold energies the total cross section exhibits quantized jumps of magnitude $4/k_0$, resulting in a plateau structure in the cross section. It would be interesting to see if these plateaus lead to a quantized conductance in the context of a transport problem. Indeed it has been shown that the singular density of states leads to non-Drude DC conductivity in the negative energy regime~\cite{brosco2016} 

In the extreme low-energy limit, the $T$-matrix becomes independent of partial wave number. This is markedly different from the conventional 2D problem where $s$-wave scattering dominates in this limit. Evidently, ultra-low-energy scattering in a Rashba system is highly anisotropic, a result which may have interesting consequences for the physics of spontaneous symmetry breaking in interacting 2D Rashba systems. 

\acknowledgements

J.H. was supported by NSERC. J.M. was supported by NSERC grant \#RGPIN-2014-4608, the Canada Research Chair Program (CRC), the Canadian Institute for Advanced Research (CIFAR), and the University of Alberta.

\appendix

\section{Rashba Green's function in position-space} \label{App:AppendixA}
Here we derive the retarded position-space Green's function. This derivation can be found in Ref.~\cite{Walls2006}, but we include it here for completeness and to standardize the notation.

We may write the Green's function as a $2\times2$ matrix in spin-space,
\begin{eqnarray}
G^+(\b{r},\b{r}';E)&=&\int\frac{d^2\b k}{4\pi^2}\frac{e^{i\b{k}\cdot(\b{r}-\b{r}')}}{(E-\frac{k^2}{2m})^2-(\lambda k)^2+i\epsilon}\nonumber\\
&&\times\begin{pmatrix}
E-\frac{k^2}{2m} & i\lambda ke^{-i\theta_\b{k}}\\
-i\lambda ke^{i\theta_\b{k}} & E-\frac{k^2}{2m}\\
\end{pmatrix}.
\end{eqnarray}

The angular integral is easily evaluated in terms of Bessel functions. For the diagonal part, one finds
\begin{eqnarray}
G^+_{\sigma\sigma}(\b{r},\b{r}';E)&=&-\frac{m}{2\pi}\int^\infty_0 dk\,J_0(k|\b{r}-\b{r}'|)\nonumber\\
&&\times\bigg(\frac{k}{k^2+2m\lambda k-2mE-i\epsilon}\nonumber\\
&&+\frac{k}{k^2-2m\lambda k-2mE-i\epsilon}\bigg).\label{eq:Greendiag1}
\end{eqnarray}
For any energy $E$, we designate the on-shell upper and lower helicity wave vectors by 
\begin{eqnarray}
k_\pm&=&\mp m\lambda+\sqrt{(m\lambda)^2+2mE}\label{eq:kHel}\\
&=&k_0(\delta\mp1).\label{eq:kHel2}
\end{eqnarray}
These determine the poles of the Green's function, which are seen from \eqref{eq:Greendiag1} by partial fraction decomposition,
\begin{eqnarray}
G^+_{\sigma\sigma}(\b{r},\b{r}';E)&=&-\frac{m}{2\pi}\int^\infty_0 dk\frac{J_0(k|\b{r}-\b{r}'|)}{k_-+k_+}\bigg(\frac{k_+}{k-k_+-i\epsilon}\nonumber\\
&&+\frac{k_-}{k+k_-+i\epsilon}+\frac{k_-}{k-k_--i\epsilon}+\frac{k_+}{k+k_++i\epsilon}\bigg).\nonumber
\end{eqnarray}

The first and last terms may be combined, as well as the second and third to give
\begin{eqnarray}
G^+_{\sigma\sigma}(\b{r},\b{r}';E)&=&-\frac{m}{\pi(k_-+k_+)}\bigg(k_+\int^\infty_0 dk \frac{kJ_0(k|\b{r}-\b{r}'|)}{k^2-(k_++i\epsilon)^2}\nonumber\\
&&+k_-\int^\infty_0 dk \frac{kJ_0(k|\b{r}-\b{r}'|)}{k^2-(k_-+i\epsilon)^2}\bigg).
\end{eqnarray}

These last integrals may be evaluated with a useful identity,
\be\label{eq:BesselIdent}
\int^\infty_0dt\,J_\nu(at)\frac{t}{t^2-z^2}=\frac{\pi i}{2}H^+_\nu(az),
\ee
valid for $a>0$, $\im z>0$.
Thus,
\begin{eqnarray}
G^+_{\sigma\sigma}(\b{r},\b{r}';E)&=&-\frac{im}{2(k_-+k_+)}\bigg(k_+H_0^+(k_+|\b{r}-\b{r}'|)\nonumber\\
&&+k_-H_0^+(k_-|\b{r}-\b{r}'|)\bigg).\label{eq:GDiag}
\end{eqnarray}

Next we evaluate the off-diagonal components. The angular integral again gives a Bessel function
\be
G^+_{\sigma\sigma'}(\b{r},\b{r}';E)=\mp\frac{1}{2\pi}\int^\infty_0dk \frac{\lambda k^2J_1(k|\b{r}-\b{r}'|)e^{\mp i\theta_{\b{r}-\b{r}'}}}{(E-\frac{k^2}{2m})^2-(\lambda k)^2+i\epsilon}.
\ee
Here the top sign is for $\sigma=\uparrow, \sigma'=\downarrow$, and the bottom is for  $\sigma=\downarrow, \sigma'=\uparrow$. Proceeding with the radial integral as before, we obtain
\begin{eqnarray}
G^+_{\sigma\sigma'}(\b{r},\b{r}';E)&=&\pm\frac{m}{2\pi}e^{\mp i\theta_{\b{r}-\b{r}'}}\int^\infty_0 dk\, k\frac{J_1(k|\b{r}-\b{r}'|)}{k_-+k_+}\nonumber\\
&&\times\bigg(\frac{1}{k-k_+-i\epsilon}-\frac{1}{k+k_-+i\epsilon}\nonumber\\
&&-\frac{1}{k-k_--i\epsilon}+\frac{1}{k+k_++i\epsilon}\bigg)\label{eq:Goff1}.
\end{eqnarray}

Both Bessel and Hankel functions satisfy the differential relation
\be
\frac{\partial}{\partial a}f_0(ax)=-xf_1(ax),
\ee
so upon combining the first and last terms as well as the second and third terms in \eqref{eq:Goff1}, we may write
\begin{eqnarray}
G^+_{\sigma\sigma'}(\b{r},\b{r}';E)&=&\mp\frac{m}{\pi(k_++k_-)}e^{\mp i\theta_{\b{r}-\b{r}'}}\nonumber\\
&&\times\frac{\partial}{\partial |\b{r}-\b{r}'|}\int^\infty_0dk\, J_0(k|\b{r}-\b{r}'|)\nonumber\\
&&\times\bigg(\frac{-k}{k^2-(k_-+i\epsilon)^2}+\frac{k}{k^2-(k_++i\epsilon)^2}\bigg).\nonumber\\
\end{eqnarray}
Using \eqref{eq:BesselIdent}, we arrive at
\begin{eqnarray}
G^+_{\sigma\sigma'}(\b{r},\b{r}';E)&=&\mp\frac{im}{2(k_-+k_+)}\bigg(k_-H_1^+(k_-|\b{r}-\b{r}'|)\nonumber\\
&&-k_+H_1^+(k_+|\b{r}-\b{r}'|)\bigg)e^{\mp i\theta_{\b{r}-\b{r}'}}.\label{eq:GOff}
\end{eqnarray}

\section{Alternative derivation of $S$ and $T$ matrix relation} \label{App:AppendixB}
We consider an alternative derivation of \eqref{eq:TS} starting from \eqref{eq:TS2}, which will also be valid for any circularly symmetric Rashba scattering problem below the Dirac point. Any negative-energy state in this system is completely characterized by three quantum numbers: energy, angular momentum, and channel index $s_\mu=\text{sgn}(k_\mu-k_0)$;
\be
|\psi\rangle=|E,l,s_\mu\rangle.
\ee
Since these are eigenstates of the unperturbed Hamiltonian, we must have
\begin{eqnarray}
0&=&(H_0-E)|E,l,s_\mu\rangle\nonumber\\
&=&\bigg(\frac{k^2}{2m}-\lambda k-E\bigg)\langle\b{k},-|E,l,s_\mu\rangle,
\end{eqnarray}
where we have used the fact that $H_0$ is diagonal in the helicity basis $|\b{k},\pm\rangle$. The overlap above is nontrivial only if 
\be
\langle\b{k},-|E,l,s_\mu\rangle\propto\delta\bigg(\frac{k^2}{2m}-\lambda k-E\bigg).
\ee
The constant of proportionality is chosen to satisfy the orthonormality conditions
\begin{eqnarray}
\langle E',l',s_\nu |E,l,s_\mu\rangle&=&\delta_{ll'}\delta_{\mu\nu}\delta(E-E'),\\
\langle\b{k}',-|\b{k},-\rangle&=&(2\pi)^2\delta(\b{k}-\b{k}').
\end{eqnarray}
One can check that the appropriate change of basis is given by
\be
\langle\b{k},-|E,l,s_\mu\rangle=\sqrt{\frac{2\pi|k_0-k|}{mk}}e^{il\theta_\b{k}}\delta_{s_\mu,s(k)}\delta\bigg(\frac{k^2}{2m}-\lambda k-E\bigg),
\ee
where $s(k)\equiv\text{sgn}(k-k_0)$.

This conversion allows us to write the momentum-space $T$-matrix starting with the $S$-matrix in the $E,l,s_\mu$ basis. From \eqref{eq:TS2},
\begin{eqnarray}
T^{\b{kk}'}_{--}\delta(E(k)-E'(k'))&=&\frac{i}{2\pi}(S-\mathbb{I})_{\b{kk}'}\\
&=&i\sum_{s_\nu,s_\rho}\sum_{l=-\infty}^{\infty}\int dE\sqrt{\frac{|k_0-k|}{mk}}\nonumber\\
&&\times\delta\bigg(\frac{k^2}{2m}-\lambda k-E\bigg)e^{il\theta_\b{k}}\delta_{s(k),s_\nu}\nonumber\\
&&\times(S^l_{\nu\rho}(E)-\delta_{\nu\rho})\sqrt{\frac{|k_0-k'|}{mk'}}\nonumber\\
&&\times\delta\bigg(\frac{k'^2}{2m}-\lambda k'-E\bigg)e^{-il\theta_{\b{k}'}}\delta_{s_\rho,s(k')},\nonumber\\
\end{eqnarray}
where we used the fact that angular momentum conservation and elastic scattering guarantee that the $S$-matrix is diagonal in $l$ and $E$. Thus we finally have
\begin{eqnarray}
T^{\b{kk}'}_{--}(E)&=&\frac{i}{m}\sqrt{\frac{|k_0-k||k_0-k'|}{kk'}}\sum_{l=-\infty}^\infty e^{il\theta}\nonumber\\
&&\times\bigg(S^l_{s(k)s(k')}-\delta_{s(k)s(k')}\bigg).
\end{eqnarray}
Letting $k=k_\nu$, $k'=k_\mu$, and noting that $|k_0-k_\nu|=|k_0-k_\mu|=k_0\delta$, we recover \eqref{eq:TS}.

\section{Derivation of the low-energy Rashba $T$-matrix}\label{app:RashbaT}
We start with the proof of \eqref{eq:Il2}. Beginning from \eqref{eq:Il}, we proceed by making the following substitutions: first, recall that $q=k_0(1+\epsilon)$, so that
\begin{eqnarray}
I^l_-
&=&\frac{m}{2\pi}\int^\Lambda_{-\Lambda}d\epsilon\frac{(1+\epsilon)[V^l(k_0,q)+V^{l+1}(k_0,q)]}{\delta^2-\epsilon^2+i\eta}.
\end{eqnarray}
Then, let $x=\delta^2-\epsilon^2$. This requires splitting the integration region into $\epsilon=\sqrt{\delta^2-x}>0$ and $\epsilon=-\sqrt{\delta^2-x}<0$:
\begin{eqnarray}
I^l_-&=&\frac{m}{4\pi}\int^{\delta^2}_{\delta^2-\Lambda^2}dx\frac{(1+\sqrt{\delta^2-x})[V^l_++V^{l+1}_+]}{(x+i\eta)\sqrt{\delta^2-x} }\nonumber\\
&&+\frac{m}{4\pi}\int^{\delta^2}_{\delta^2-\Lambda^2}dx\frac{(1-\sqrt{\delta^2-x})[V^l_-+V^{l+1}_-]}{(x+i\eta)\sqrt{\delta^2-x}},\nonumber\\\label{eq:Ilm}
\end{eqnarray}
where we have defined $V^l_{\pm}\equiv V^l(k_0,k_0(1\pm\sqrt{\delta^2-x}))$. The imaginary part of these integrals is readily found,
\begin{eqnarray}
\im I^l_-&=&-\frac{m}{4\delta}\bigg((1+\delta)[V^l_++V^{l+1}_+]\nonumber\\
&&\left.+(1-\delta)[V^l_-+V^{l+1}_-]\bigg)\right|_{x=0}.
\end{eqnarray}
Expanding about $\delta=0$ gives
\be
\im I^l_-=-\frac{m}{2\delta}[V^l(k_0,k_0)+V^{l+1}(k_0,k_0)]+\mathcal{O}(\delta)\label{eq:im}.
\ee
Note that the interference between virtual states with $q<k_0$ and $q>k_0$ causes the cancellation of the $\mathcal{O}(1)$ term in \eqref{eq:im}.

The only thing left is to consider the real (or principal) part of \eqref{eq:Ilm}. We will show that this term gives the cutoff-dependent corrections to the $T$-matrix \eqref{eq:TfirstOrder}. The trick is to isolate the momentum dependence of the potential by making use of the following multiplication theorem for Bessel functions~\cite{NIST:DLMF},
\be
J_\nu(\lambda z)=\lambda^\nu\sum^\infty_{k=0}\frac{(-1)^k(\lambda^2-1)^k(z/2)^k}{k!}J_{\nu+k}(z).
\ee
We may apply this to \eqref{eq:Vl} to first isolate the angular dependence,
\begin{eqnarray}
V^l(k_0,q)&=&\sum_{k=0}^\infty\frac{1}{k!}\int^{2\pi}_0\frac{d\theta}{2\pi}2^{-k}(e^{i\theta}+e^{-i\theta})^ke^{il\theta}\nonumber\\
&&\times\int^\infty_0dr\,rV(r)\left(\frac{k_0r}{\sqrt{2}}\sqrt{1+\epsilon}\right)^k\nonumber\\
&&\times J_k(\sqrt{2}k_0r\sqrt{1+\epsilon}),\nonumber\\
\end{eqnarray}
where convergence of the infinite series allows us to take it outside the integral. The $\theta$ integral is easily evaluated with the binomial theorem,
\begin{eqnarray}
V^l(k_0,q)&=&\sum_{k=0}^\infty\frac{2^{-k}}{k!}\sum_{n=0}^k {k\choose n}\delta_{n,\frac{k+l}{2}}\int^\infty_0dr\,rV(r)\nonumber\\
&&\times\left(\frac{k_0r}{\sqrt{2}}\sqrt{1+\epsilon}\right)^kJ_k(\sqrt{2}k_0r\sqrt{1+\epsilon}).
\end{eqnarray}
Making a change of summation variables $k\rightarrow|l|+2k$, and applying the same multiplication theorem to the remaining Bessel function, we get
\begin{eqnarray}
V^l(k_0,q)&=&\sum^\infty_{k=0}\frac{2^{-\frac{3}{2}(|l|+2k)}}{(k+|l|)!k!}(1+\epsilon)^{|l|+2k}\sum_{n=0}^\infty\frac{(-1)^n\epsilon^n}{n!}\nonumber\\
&&\times\int^\infty_0dr\,rV(r)\frac{(k_0r)^{|l|+2k+n}}{2^{n/2}}J_{|l|+2k+n}(\sqrt{2}k_0r),\nonumber\\
\end{eqnarray}
or equivalently,
\begin{eqnarray}
V^l_{\pm}&=&\sum_{n=0}^\infty\sum_{k=0}^\infty f_{nk|l|}\left(1\pm\sqrt{\delta^2-x}\right)^{|l|+2k}\left(\pm\sqrt{\delta^2-x}\right)^n,\nonumber\\
\end{eqnarray}
where he have defined
\begin{eqnarray}
f_{nk|l|}&\equiv&\frac{2^{-\frac{3}{2}(|l|+2k+n/3)}}{(k+|l|)!k!}\frac{(-1)^n}{n!}\nonumber\\
&&\times\int^\infty_0 dr\,rV(r)(k_0r)^{|l|+2k+n}J_{|l|+2k+n}(\sqrt{2}k_0r).\nonumber\\
\end{eqnarray}
Inserting this into \eqref{eq:Ilm} gives
\begin{eqnarray}
\re I_-^l&=&\frac{m}{4\pi}\sum_{n=0}^\infty\sum_{k=0}^\infty f_{nk|l|}\mathcal{P}\int^{\delta^2}_{\delta^2-\Lambda^2}\frac{dx}{x\sqrt{\delta^2-x}}\nonumber\\
&&\times\bigg[(\sqrt{\delta^2-x})^n(1+\sqrt{\delta^2-x})^{|l|+2k+1}\nonumber\\
&&+(-1)^n(\sqrt{\delta^2-x})^n(1-\sqrt{\delta^2-x})^{|l|+2k+1}\bigg]\nonumber\\
&&+(l\rightarrow l+1),
\end{eqnarray}
where $\mathcal{P}$ denotes the principal value and the last line means we add the previous lines with $l$ replaced by $l+1$. 
These integrals may be solved exactly, but here we only consider the lowest order terms in the small parameter $\sqrt{\delta^2-x}<\Lambda\ll1$. The square brackets above may be expanded in this parameter to give $\delta_{n,0}+\mathcal{O}(\delta^2-x)$. The fact that no terms of order $\sqrt{\delta^2-x}$ appear in these brackets is due to the interference between $q<k_0$ and $q>k_0$ states. It is these absent terms that would have yielded the logarithmic dependence $\ln(\delta/\Lambda)$ were this conventional 2D scattering. With this approximation, the integrals are readily evaluated as
\begin{eqnarray}
\re I_-^l&\approx&\frac{m}{2\pi}\sum_{k=0}^\infty\frac{2}{\Lambda}\left( f_{0k|l|}+ f_{0k|l+1|}\right),\nonumber\\
\end{eqnarray}
where the terms neglected in this approximation are $\mathcal{O}(\Lambda)$.
Noting that 
\be
\sum_{k=0}^\infty f_{0k|l|}=V^l(k_0,k_0),\label{eq:f0kl}
\ee
we summarize this result as
\be
\re I^l_-\approx\frac{m}{\pi\Lambda}\left(V^l(k_0,k_0)+V^{l+1}(k_0,k_0)\right).
\ee
Thus we can approximate the $T$-matrix by
\be
T^l_{--}\approx\frac{\frac{1}{2}[V^l(k_0,k_0)+V^{l+1}(k_0,k_0)]}{1+\frac{m}{2}(\frac{i}{\delta}+\frac{2}{\pi\Lambda})[V^l(k_0,k_0)+V^{l+1}(k_0,k_0)]},
\ee
which is linear in $\delta$ to leading order, with a $\mathcal{O}(\delta^2)$ correction to subleading order. The correction to the approximation $I^l_-$ contributes to $\mathcal{O}(\delta^3)$. Now one might make the following objection. The first approximation we made in Sec.~\ref{sec:RashbaT} was $V_{ji}(\b{k}_\nu,\b{k}_\mu)=V_{ji}(k_0\hat{\b{k}}_\nu,k_0\hat{\b{k}}_\mu)+\mathcal{O}(\delta)$ and $V_{ji}(\b{k}_\nu,\b{q})=V_{ji}(k_0\hat{\b{k}}_\nu,\b{q})+\mathcal{O}(\delta)$. A glance at the Born series \eqref{eq:TBorn} suggests that there will be corrections to the $T$-matrix of order $\delta$ as well. However, this is not the case since we are focusing on the nonperturbative regime of the Born series. To be specific, let $T'$ be the corrections in the $T$-matrix due to the $\mathcal{O}(\delta)$ corrections in the potential,
\begin{equation}
T=T^0+T',\hspace{5mm}
V=V^0+\delta V',
\end{equation}
where $T^0$ is the leading-order $T$-matrix approximation that was just derived. In terms of operators, the Born series now reads
\be
T'=\delta V'(1+G^+T^0)+(V^0+\delta V')G^+T'.
\ee
We can then apply the same arguments as before. In terms of helicity and momentum-space components, the right-hand side of this equation is independent of $\b{k}_\nu$, and so $T'$ is as well. Expanding in partial waves and ignoring interband scattering gives
\begin{eqnarray}
\sum_{l=-\infty}^\infty T'^{l}_{--}(k_\mu) e^{il\theta}&=&\sum_{l=-\infty}^\infty\frac{\delta}{2} e^{il\theta}\bigg\{[V'^l(k_0,k_0)\nonumber\\
&&+V'^{l+1}(k_0,k_0)]+2I^l_-T^{0l}_{--}(k_\mu)\bigg\}\nonumber\\
&+& \sum_{l=-\infty}^\infty e^{il\theta}\bigg\{ I^l_-+\delta\int\frac{dq}{2\pi}q\bigg(V'^l(k_0,q)\nonumber\\
&&+ V'^{l+1}(k_0,q)\bigg)G^+_{--}(q)\bigg\}T'^l_{--}(k_\mu).\nonumber\\
\end{eqnarray}
Defining
\begin{equation}
I'^l_-\equiv\int\frac{dq\,q}{4\pi}(V'^l(k_0,q)+V'^{l+1}(k_0,q))G^+_{--}(q),
\end{equation}
and solving for $T'^l_{--}$ gives
\be
T'^l_{--}=\delta\frac{\frac{1}{2}(V'^l(k_0,k_0)+V'^{l+1}(k_0,k_0))+T^{0l}_{--}I^l_-}{1-I^l_--\delta I'^l_-}.
\ee
However, we know that to lowest order, $I^l_-\sim 1/\delta$ and $T^{0l}_{--}\sim\delta$, so the numerator above is constant. Meanwhile, the derivation of $I^l_-$ did not depend on the details of the potential components $V^l(k_0,q)$ and so applies equally to $I'^l_-$, giving $I'^l_-\sim 1/\delta$. The denominator is therefore dominated by the $I^l_-$ term so that
\be
T'^l_{--}\sim\delta^2.
\ee
Hence our approximation $T^{0l}_{--}$ is valid to order $\delta^2$.

\bibliography{UniversalRashba}

\end{document}